\documentclass[cameraready]{Interspeech}

\title{Too Good to Be True: A Study on Modern Automatic Speech Recognition \\for the Evaluation of Speech Enhancement}

\author[orcid=0009-0001-3338-7913]{Danilo}{{de Oliveira}}
\author[orcid=0000-0002-8974-9127]{Tal}{Peer}
\author[orcid=0000-0002-8678-4699]{Timo}{Gerkmann}

\address{University of Hamburg, Signal Processing Group}

\email{danilo.oliveira@uni-hamburg.de}


\usepackage{comment}
\usepackage{booktabs}
\usepackage{multirow}
\usepackage{microtype}
\usepackage[nolist]{acronym}
\usepackage{adjustbox}
\usepackage{xcolor}
\usepackage{colortbl}

\definecolor{rankI}{RGB}{26,150,65}
\definecolor{rankII}{RGB}{166,217,106}
\definecolor{rankIII}{RGB}{217,239,139}
\definecolor{rankIV}{RGB}{255,255,191}
\definecolor{rankV}{RGB}{252,174,97}
\definecolor{rankVI}{RGB}{240,95,95}
\definecolor{cleanblue}{RGB}{173,216,230}

\begin{acronym}

\acro{acr}[ACR]{Absolute Category Rating}
\acro{asr}[ASR]{automatic speech recognition}
\acro{ast}[AST]{automatic speech translation}
\acro{cfg}[CFG]{classifier-free guidance}
\acro{ci}[CI]{confidence interval}
\acro{ctc}[CTC]{connectionist temporal classification}
\acro{dnn}[DNN]{deep neural network}
\acro{estoi}[ESTOI]{extended short-term objective intelligibility}
\acro{gan}[GAN]{generative adversarial network}
\acro{gmm}[GMM]{Gaussian mixture model}
\acro{ha}[HA]{hearing aid}
\acro{hmm}[HMM]{hidden Markov model}
\acro{istft}[iSTFT]{inverse short-time Fourier transform}
\acro{l1}[L1]{first language}
\acro{l2}[L2]{second language}
\acro{lm}[LM]{language model}
\acro{llm}[LLM]{large language model}
\acro{lpd}[LPD]{Levenshtein phoneme distance}
\acro{lrs}[LRS]{Lip Reading Sentences}
\acro{lsd}[LSD]{log spectral distance}
\acro{lstm}[LSTM]{long short-term memory}
\acro{mac}[MAC]{multiply-and-accumulate}
\acro{mer}[MER]{match error rate}
\acro{mos}[MOS]{mean opinion score}
\acro{mse}[MSE]{mean squared error}
\acro{oa}[OA]{observation adding}
\acro{ode}[ODE]{ordinary differential equation}
\acro{per}[PER]{phoneme error rate}
\acro{pesq}[PESQ]{Perceptual Evaluation of Speech Quality}
\acro{polqa}[POLQA]{Perceptual Objective Listening Quality Analysis}
\acro{sde}[SDE]{stochastic differential equation}
\acro{sdr}[SDR]{signal-to-distortion ratio}
\acro{se}[SE]{speech enhancement}
\acro{sip}[SIP]{speech intelligibility prediction}
\acro{sisdr}[SI-SDR]{scale-invariant signal to distortion ratio}
\acro{slm}[sLM]{speech language model}
\acro{snr}[SNR]{signal-to-noise ratio}
\acro{sota}[SotA]{state-of-the-art}
\acro{sqa}[SQA]{speech quality assessment}
\acro{srt}[SRT]{speech reception threshold}
\acro{ssl}[SSL]{self-supervised learning}
\acro{stft}[STFT]{short-time Fourier transform}
\acro{stt}[STT]{speech-to-text}
\acro{tf}[T-F]{time-frequency}
\acro{tts}[TTS]{text-to-speech}
\acro{uhh}[UHH]{University of Hamburg}
\acro{vae}[VAE]{variational auto-encoder}
\acro{vbdmd}[VB-DMD]{Voicebank-DEMAND}
\acro{vsr}[VSR]{video speech recognition}
\acro{wacc}[WAcc]{word accuracy}
\acro{wer}[WER]{word error rate}
\acro{wil}[WIL]{word information lost}
\acro{wip}[WIP]{word information preserved}
\acro{wp}[WP]{work package}

\end{acronym}

\begin{document}

\maketitle

\begin{abstract}
    Speech enhancement (SE) systems are typically evaluated using a variety of instrumental metrics. The use of automatic speech recognition (ASR) systems to evaluate SE performance is common in literature, usually in terms of word error rate (WER). However, WER scores depend heavily on the choice of ASR system and text normalization pipeline. In this paper, we investigate how modern ASR models correlate with human recognition of enhanced speech. A listening experiment reveals that modern ASR models with large-scale noisy training and embedded language models correlate more with human WER than simpler ones, with a transducer model providing the most reliable transcriptions. Nevertheless, we also show that these models' robustness to noise and use of context can be uninformative to an acoustics-focused evaluation of enhancement performance.
\end{abstract}

\section{Introduction}

The task of \ac{se} generally aims at improving the quality and intelligibility of degraded speech recordings. There exists a multitude of instrumental metrics to evaluate \ac{se} performance without having to resort to human-generated scores, which are costly to collect, time- and effort-wise. As a way of avoiding biases towards a specific measure or a type of \ac{se} model, it is a good practice to use not only one but an ensemble of evaluation metrics to paint a complete picture of a system's performance~\cite{deoliveira2024pesqetarian}. It is not rare to find \ac{wer} among the metrics of choice. For example, The 2022 edition of the DNS Challenge used the performance of \ac{asr} on denoised speech as one of the criteria to rank submissions~\cite{dubey2022deepnoise}. Nevertheless, the purpose of using \ac{wer} as a metric is not always clearly expressed, nor is the choice of \ac{asr} model for transcriptions.

In the context of \ac{se} evaluation, the quality dimension tends to draw most attention, typically reported in terms of intrusive (reference-based) measures such as PESQ~\cite{rix2001pesq} and SI-SDR~\cite{leroux2019sdr}, and non-intrusive (reference-free) metrics, like NISQA~\cite{mittag2021nisqa}, DNSMOS~\cite{reddy2022dnsmosp835} and UTMOS~\cite{saeki2022utmos}.
The intelligibility of enhanced speech is often predicted instrumentally using STOI~\cite{taal2011algorithm} and its extension ESTOI~\cite{jensen2016algorithm}.
Both STOI and ESTOI are signal-intrusive and do not require a reference text. However, their relation with human intelligibility ratings is non-linear and dataset-dependent~\cite{taal2011algorithm,jensen2016algorithm}, rendering the interpretation of relative differences in (E)STOI values and linear correlation coefficients difficult.

There is extensive literature showing that \ac{asr} models can be modified to predict speech intelligibility. Karbasi and Kolosssa~\cite{karbasi2022asrbased} reviewed \ac{asr} methods for \ac{sip}, either using their outputs or their internals. A prominent example of the first category is the the FADE framework by Sch\"adler et al.~\cite{schaedler2015matrix}, which predicts the \ac{srt} by training multiple \ac{asr} systems.
Examples of the latter line of work are the methods from Karbasi et al.~\cite{karbasi2020nonintrusive} and Arai et al.~\cite{arai2020predicting}. Most \ac{asr}-based \ac{sip} make use of ``traditional'' hybrid models using \acp{gmm} or \acp{dnn} to model acoustic features and \acp{hmm} for temporal modeling of acoustic units; meanwhile, the realm of modern end-to-end models remains underexplored.

In a broader view of \ac{se} evaluation, Siddiqui et al.~\cite{siddiqui2020using} investigated using \ac{asr} to compare \ac{dnn}-based \ac{se} methods. Nevertheless, only one \ac{asr} model was used, chosen based on its poor noise robustness, and most importantly, only correlations with instrumental metrics and \ac{snr} were analyzed, not human scores. De Oliveira et al.~\cite{deoliveira2025arethese} showed that perplexity scores of a \ac{lm} on \ac{asr} transcriptions can be used to identify clean, noisy and even gibberish speech in a reference-free setting.
Looking into the causes for degradation of \ac{asr} performance, Iwamoto et al.~\cite{iwamoto2022howbad} decomposed \ac{se} errors into noise and artifact components, and showed that the latter one is the main cause of \ac{asr} performance degradation. The authors proposed the \ac{oa} post-processing method for improving downstream \ac{asr} performance via interpolation of noisy and enhanced signals. Further work showed that this is also the case for human intelligibility~\cite{araki2023impact} and focused on the joint training of \ac{se} and \ac{asr} from this decomposition perspective~\cite{iwamoto2024howdoes}.

Although \ac{wer} is popular metric, we argue that its use is not free of caveats. The calculation of \ac{wer} involves several choices that can considerably affect the result. From the most obvious aspect: the choice of \ac{asr} model; to the less apparent ones, such as the text normalization pipeline, important details can be easily overlooked and are often not disclosed. In this paper, we investigate how different modern \ac{asr} models, with varied losses, sizes and training data, behave when faced with noise and \ac{dnn}-based \ac{se} artifacts. Our study highlights the importance of the choice of \ac{asr} system and the clear disclosure of the chosen setup, all of which help in the understanding of what is really being evaluated when reporting \ac{wer} values.

\section{Method}

\subsection{Word error rate}

\Ac{wer} is the most common metric to evaluate the performance of \ac{asr} systems. It is derived from the Levenshtein (edit) distance, which measures the difference between sequences~\cite{levenshtein1966binary}. More precisely, it measures the minimum number of edits required to change one sequence into another. 

\Ac{wer} is calculated as
\begin{equation}
    \mathrm{WER} = \frac{S + D + I}{S + D + C},
    \label{eq:wer}
\end{equation}
where $S$, $D$, $I$ and $C$ are the number of substitutions, deletions, insertions and correct words, respectively. The denominator $S + D + C$ corresponds to the total number of words in the reference text. Being a measure of error, lower values mean better recognition performance. To obtain a measure where higher is better, \ac{wer} can be inverted into \ac{wacc}:
\begin{equation}
    \mathrm{WAcc} = 1 - \mathrm{WER} = \frac{C - I}{S + D + C}.
\end{equation}

\Ac{wer} and \ac{wacc} values are most commonly reported in percentage format. Note that when the number of insertions is larger than the number of correct words, the \ac{wacc} is negative (i.e.: \ac{wer} greater than 100\%). 

\subsection{Automatic speech recognition models}
\label{subsec:asr}

Similarly to other areas in speech and language technology, the field of \ac{asr} has benefited greatly from the advances in \acp{dnn} research. From substituting acoustic \ac{gmm} models working in tandem with \acp{hmm}, to single-handedly executing the task end-to-end, \acp{dnn} are ubiquitous in modern \ac{asr}. Common approaches are encoder-only models trained with the \ac{ctc} loss~\cite{graves2006connectionist}, encoder-decoder transducer models~\cite{graves2012sequence} and encoder-decoder attention models~\cite{chorowski2015attentionbased, chan2016las}. While the \ac{ctc} training objective has the great advantage of dispensing with pre-computed audio-text alignments, it does not model the dependencies between
the text outputs. As a consequence, \ac{ctc} models are prone to linguistic mistakes and often employ external \acp{lm} in the decoding. Transducer models address this by incorporating a decoder in the training, joining encoder and decoder representations to predict tokens. Attention-based models lose the monotonic alignment constraint, enabling \ac{ast}~\cite{berard2018endtoend}.

Historically, \ac{asr} research had been heavily concentrated on improving the \ac{sota} in specific benchmarks, the most well-known being the LibriSpeech dataset~\cite{panayotov2015librispeech}. Also in a similar fashion to \acp{llm}, \ac{asr} systems have reaped big rewards from large-scale training data and computing power, from matching human transcription performance~\cite{amodei2016deepspeech2}, to significantly surpassing it in a matched setting~\cite{baevski2020wav2vec2}, to closing the gap with human robustness in mismatched test settings~\cite{radford2023robust}. This chronology evidences the fact that the objective of \ac{asr} is not to mimic human transcription performance, but to obtain the best performance possible~\cite{karbasi2020nonintrusive}.

Below we list the models we considered in this study. Focusing on ease of use and speed required for metrics, we selected models with less than 1~B parameters. All transcriptions were generated via greedy decoding, without external \acp{lm}.

\noindent\textbf{QuartzNet~\cite{kriman2020quartznet}:} A compact convolutional model trained with the \ac{ctc} loss. Without an external \ac{lm}, QuartzNet-15x5 (18.9~M parameters) obtains 96.1\% \ac{wacc} on LibriSpeech test \emph{clean}.

\noindent\textbf{wav2vec2~\cite{baevski2020wav2vec2}:} A \ac{ssl}-based model pre-trained with a contrastive learning objective to identify masked quantized representations, and then fine-tuned for \ac{asr} in a supervised fashion with the \ac{ctc} loss. Without an external \ac{lm}, wav2vec2 LARGE LV-60k (317~M parameters) obtains 97.8\% \ac{wacc} on LibriSpeech test clean. 

\noindent\textbf{Parakeet TDT:} A Transducer model with a FastConformer encoder~\cite{rekesh2023fast} pre-trained with the wav2vec2 \ac{ssl} objective and a second training stage with the Token-and-Duration Transducer (TDT) architecture~\cite{xu2023efficient}, which allows for faster decoding. Parakeet TDT v2 (600~M parameters) was trained on a large-scale dataset of $\sim$120,000 hours of English speech and obtains 98.31\% \ac{wacc} on LibriSpeech test clean.

\noindent\textbf{Whisper~\cite{radford2023robust}:} An attention encoder-decoder model based on the original Transformer architecture~\cite{vaswani2017attention}. It is trained for next token prediction, and can leverage multilingual large scale data for multi-task translation and transcription. Multiple model sizes were trained, and pruned (Turbo) and distilled (Distil-Whisper) versions~\cite{gandhi2023distilwhisper} of the Large model also exist. Due to the large-scale weakly supervised training and the large decoder possessing general knowledge of language, Whisper can output text that was not uttered in the audio input (so-called \textit{hallucinations})~\cite{frieske2024hallucinations, koenecke2024careless, baranski2025investigation}. Whisper is generally available in English-only or multilingual versions, except for Whisper Large and its variants, which are multilingual-only. When using any multilingual version, we specify the language as English and task as transcription, to avoid language identification issues at challenging conditions.

\subsection{Speech enhancement models}
\label{subsec:se}

All \ac{se} models used in this paper were trained on the VB-DMD dataset~\cite{valentini2016investigating}. As evidenced in previous works, predictive (mapping-based) and generative models produce different types of artifacts, which are penalized differently depending on the metric~\cite{deoliveira2023behavior, pirklbauer2023evaluation}. We made our selection aiming at contemplating both paradigms, as well a hybrid methods that present characteristics from both.

\noindent\textbf{SGMSE+~\cite{richter2023speech}:} A generative diffusion-based model. It starts the generation process from the noisy input and Gaussian noise, and gradually and iteratively the noise is removed. The \ac{dnn}'s output corresponds to a term in a \ac{sde}, which is solved with multiple solver steps. The so-called score model backbone is the NCSN++ architecture, a UNet with 2D ResNet blocks totaling 65~M parameters, operating here on complex-valued spectrograms.

\noindent\textbf{SB-SGMSE+ (M8)~\cite{richter2025investigating}:} The same model architecture from SGMSE+, but trained with a Schr\"odinger bridge formulation. It can start the generative process from the noisy audio input directly and additionally allows for a data prediction loss. Richter et al.~\cite{richter2025investigating} leverage this to incorporate a differentiable PESQ loss term, consequently obtaining improved perception-based scores.

\noindent\textbf{NCSN++M~\cite{lemercier2023analysing}:} A lighter version of the NCSN++ architecture, trained for complex spectral mapping with a \ac{mse} training objective, i.e., in a predictive manner. Noise level conditioning layers are removed and the number of blocks per resolution are ablated, yielding a model with roughly half the original number of parameters, namely 27.8~M.

\noindent\textbf{StoRM~\cite{lemercier2023storm}:} A cascaded, hybrid predictive / generative system. Instead of initializing the diffusion generation process from the noisy input as in SGMSE+, StoRM starts from the outputs of NCSN++M. This means the noisy audio is first enhanced predictively, then generatively. This hybrid approach helps reduce the occurrence of generative artifacts such as breathing vocalizations and also reduces the number of generative steps required for obtaining high-quality enhanced speech.

\noindent\textbf{SE-Mamba~\cite{chao2024investigation}:} A predictive model that incorporates the state-space-based Mamba blocks~\cite{gu2024mamba} into the MP-SENet architecture~\cite{lu2023mpsenet} and a multi-task loss. It features separate magnitude and phase decoders, and a training objective including loss terms on magnitude, phase, complex spectrum and time-domain, and additionally a GAN-based PESQ loss.

\begin{table*}[t]
    \centering
    \caption{Speech enhancement methods (columns) evaluated on the listening experiment subset of EARS-WHAM test, according to word recognition (clipped WAcc), phoneme recognition (LPS), intelligibility (ESTOI) and quality (POLQA, SCOREQ) metrics. 
    }
    \label{tab:main}
\begin{adjustbox}{width=0.8\textwidth}
\begin{tabular}{clccccccc}
    \toprule
     & & & & \multicolumn{5}{c}{SE Model} \\
    \cmidrule{5-9}
     & & & & \multicolumn{2}{c}{Predictive} & \multicolumn{2}{c}{Hybrid} & Generative \\
    \cmidrule(lr){5-6}\cmidrule(lr){7-8}\cmidrule(lr){9-9}
    \multicolumn{2}{l}{Metric / ASR Model} & Clean & Noisy & SE-Mamba & NCSN++M & StoRM & SB-SGMSE+ & SGMSE+ \\
    \midrule
    \parbox[c]{2mm}{\multirow{8}{*}{\rotatebox[origin=c]{90}{WAcc [\%]}}}
     & Human
     & \cellcolor{cleanblue}95.1
     & \cellcolor{rankI}85.6
     & \cellcolor{rankIII}77.7
     & \cellcolor{rankII}81.2
     & \cellcolor{rankIV}76.7
     & \cellcolor{rankV}76.1
     & \cellcolor{rankVI}69.0 \\
    \cmidrule(l){2-9}
     & QuartzNet 15x5
     & \cellcolor{cleanblue}94.6
     & \cellcolor{rankVI}58.2
     & \cellcolor{rankI}72.7
     & \cellcolor{rankII}71.2
     & \cellcolor{rankIII}66.2
     & \cellcolor{rankIV}62.3
     & \cellcolor{rankV}59.7 \\
     & wav2vec2 (L)
     & \cellcolor{cleanblue}96.1
     & \cellcolor{rankV}70.2
     & \cellcolor{rankII}76.5
     & \cellcolor{rankI}81.1
     & \cellcolor{rankIII}76.3
     & \cellcolor{rankIV}74.3
     & \cellcolor{rankVI}68.5 \\
     & Parakeet TDT v2
     & \cellcolor{cleanblue}97.0
     & \cellcolor{rankI}95.0
     & \cellcolor{rankIII}87.2
     & \cellcolor{rankII}89.8
     & \cellcolor{rankIV}85.6
     & \cellcolor{rankV}85.2
     & \cellcolor{rankVI}73.4 \\
     & Whisper Base (En)
     & \cellcolor{cleanblue}97.0
     & \cellcolor{rankI}85.9
     & \cellcolor{rankIII}81.1
     & \cellcolor{rankII}82.1
     & \cellcolor{rankIV}80.2
     & \cellcolor{rankV}77.7
     & \cellcolor{rankVI}68.7 \\
     & Whisper Base
     & \cellcolor{cleanblue}97.4
     & \cellcolor{rankI}85.4
     & \cellcolor{rankIII}81.4
     & \cellcolor{rankII}83.0
     & \cellcolor{rankIV}78.2
     & \cellcolor{rankV}72.5
     & \cellcolor{rankVI}64.4 \\
     & Whisper Large v3 Turbo
     & \cellcolor{cleanblue}98.1
     & \cellcolor{rankI}94.1
     & \cellcolor{rankIII}86.7
     & \cellcolor{rankII}91.4
     & \cellcolor{rankIV}85.8
     & \cellcolor{rankV}84.7
     & \cellcolor{rankVI}77.9 \\
     & Distil-Whisper Large v3
     & \cellcolor{cleanblue}98.1
     & \cellcolor{rankI}93.4
     & \cellcolor{rankIII}84.8
     & \cellcolor{rankII}89.4
     & \cellcolor{rankV}82.9
     & \cellcolor{rankIV}83.7
     & \cellcolor{rankVI}73.5 \\
    \midrule\midrule
     & LPS [\%]
     & {---}
     & \cellcolor{rankVI}59.2
     & \cellcolor{rankII}79.1
     & \cellcolor{rankI}82.7
     & \cellcolor{rankIV}75.2
     & \cellcolor{rankIII}76.6
     & \cellcolor{rankV}70.0 \\
     & ESTOI [\%]
     & {---}
     & \cellcolor{rankVI}56.0
     & \cellcolor{rankIV}72.1
     & \cellcolor{rankIII}73.1
     & \cellcolor{rankII}73.3
     & \cellcolor{rankI}74.2
     & \cellcolor{rankV}71.0 \\
    \midrule
     & POLQA
     & {---}
     & \cellcolor{rankVI}1.86
     & \cellcolor{rankI}2.79
     & \cellcolor{rankIV}2.38
     & \cellcolor{rankII}2.55
     & \cellcolor{rankV}2.37
     & \cellcolor{rankIII}2.41 \\
     & SCOREQ
     & \cellcolor{cleanblue}4.59
     & \cellcolor{rankVI}1.91
     & \cellcolor{rankIII}3.07
     & \cellcolor{rankV}2.80
     & \cellcolor{rankII}3.17
     & \cellcolor{rankIV}3.04
     & \cellcolor{rankI}3.46 \\
    \midrule\midrule
    \multicolumn{2}{l}{\textit{Color scale (rank)}}
     & \cellcolor{cleanblue}{---}
     & \cellcolor{rankI}\textcolor{white}{\textbf{1}}
     & \cellcolor{rankII}\textbf{2}
     & \cellcolor{rankIII}\textbf{3}
     & \cellcolor{rankIV}\textbf{4}
     & \cellcolor{rankV}\textbf{5}
     & \cellcolor{rankVI}\textcolor{white}{\textbf{6}} \\
    \bottomrule
\end{tabular}
\end{adjustbox}
\end{table*}

\subsection{Speech enhancement metrics}
We employ several other \ac{se} metrics for comparison:

\noindent\textbf{POLQA~\cite{beerends2013perceptual}:} A full reference approach to predict listening quality. It uses a psycho-acoustic model to transform reference and degraded audio into perceptually-motivated representations, and compares them. Discrepancies are mapped into penalties in the \ac{acr} scale, from 1 to 5. POLQA was standardized as ITU-T Recommendation P.863 \cite{itu-t-p-863}, superseding PESQ. We use the full-band mode of POLQA v3.

\noindent\textbf{SCOREQ~\cite{ragano2024scoreq}:} A reference-free model predicting speech quality in the \ac{acr} scale. It is a wav2vec2 model fine-tuned with a contrastive triplet regression loss that structures the embedding space based on the proximity of the corresponding \ac{mos} labels.

\noindent\textbf{ESTOI~\cite{jensen2016algorithm}:} An extension of STOI, a full-reference approach to predict intelligibility. It involves extracting normalized temporal envelopes in frequency sub-bands followed by spectral correlation coefficients. These are then averaged across time to produce an index that is expected to have a monotonic relation with speech intelligibility.

\noindent\textbf{LPS~\cite{pirklbauer2023evaluation}:} A method aiming to address the issue of phoneme confusions in generative models. A wav2vec2-based phoneme classifier~\cite{xu2022deepnoise} transcribes enhanced audio and its clean reference phonemically. These transcriptions are then used to compute the phoneme accuracy (phoneme-level equivalent of \ac{wacc}).

\subsection{Listening experiment}

In order to compare \ac{asr} performance with human transcription capabilities on speech corrupted by noise and \ac{dnn}-generated artifacts, we conducted a listening experiment with 20 participants from diverse backgrounds. The data used in the experiment was selected from the gender-varied, English-language EARS-WHAM~\cite{richter2024ears} test set, down-sampled to 16~kHz. To focus on transcribing behavior when faced with corruptions, we only considered noisy mixtures with \acp{snr} between $-2.5$ and $10$~dB. These samples were fed into the \ac{se} systems from Sec.~\ref{subsec:se}.

We selected 30 speech files for which ground-truth transcripts were available (EARS also contains freeform speech), discarding the utterances of the rainbow passage. To match typical \ac{se} evaluation settings, participants listened to complete audio files containing multiple phonetically rich sentences, albeit only loosely connected semantically with one another. The average duration was 11 seconds. Samples were distributed across participants keeping the same average input \ac{snr}, in order to balance the difficulty level. 
Each participant transcribed three files from each system, including clean and noisy. The distribution also intended to avoid the amount of sentence repetition as much as possible (one at most), so that participants would not be influenced by past samples. Participants were allowed to pause and replay the audio files, and were instructed to indicate incomprehensible speech with the tag \texttt{<UNK>}. Except where specified otherwise, for \ac{wer} / \ac{wacc} computation we normalized all text, from humans and \ac{asr} systems, expanding on the transformation \texttt{wer\_standardize} from the Python library \texttt{jiwer}. We additionally remove punctuation, expand the informal contractions \emph{gonna} and \emph{wanna}, and convert numbers into text form. On the clean data, all participants obtained \ac{wacc} higher than 90\%, with all clean transcriptions averaging 95.1\%. 

\begin{figure*}[h]
  \centering
  \includegraphics[width=\linewidth]{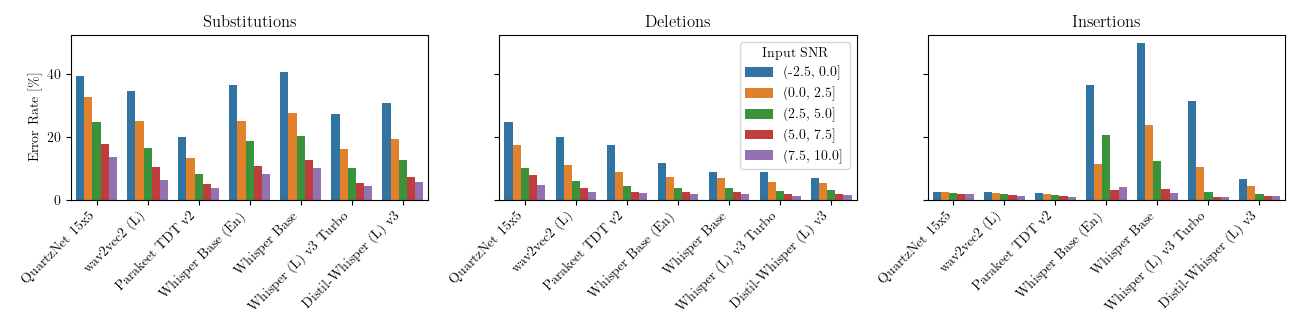}
  \vspace{-20pt}
  \caption{Decomposition of error sources for each \ac{asr} model, across all enhanced audio systems. Substitution, deletion and insertion rates, respectively, according to Equation~\ref{eq:split}. These errors are visualized across input \ac{snr}, grouped in bins at $2.5$~dB intervals.}
  \vspace{-10pt}
  \label{fig:errors}
\end{figure*}

\section{Results}

In order to facilitate the interpretation of correlations with other standard \ac{se} metrics where higher values are better, we compute the \ac{wacc} instead of the \ac{wer}. Moreover, to mitigate the effect of catastrophic failures of Whisper while sticking to the typical aggregation procedure of averaging \ac{wer} / \ac{wacc} scores, we clip \ac{wacc} by computing $\mathrm{WAcc} = \max(1 - \mathrm{WER}, 0)$. Without these measures, outliers appear which severely affect the average scores and correlations, as further discussed in Sec.~\ref{subsec:sdi}. Alternative measures could be to use the \ac{mer}~\cite{morris2004fromwer}, which is naturally constrained between 0 and 1, or to aggregate the \acp{wacc} via median instead of the mean.

Table~\ref{tab:main} shows the \ac{wacc} for each \ac{asr} system, alongside the other \ac{se} metrics. Endowed with language modeling capabilities and large-scale with noisy data, Parakeet and Whisper outperformed the human annotators across all settings. Nevertheless, their trends were also the closest to that of the human subjects, as shown in Table~\ref{tab:corr}. Due to the limited number of \ac{se} systems, for system-level correlations we computed \acp{ci} via 5,000 bootstrapping samples. The CTC models suffered more with the degraded conditions, presenting more variation in the range of scores. For Parakeet and Whisper, as well as humans, the enhancement of speech is counterproductive to speech recognition, as the \ac{wacc} is higher for the noisy subset than for any enhanced subset. This confirms findings from previous works on \ac{oa}~\cite{iwamoto2022howbad, araki2023impact}, and is likely due to the presence of noisy audio in the large-scale training, but lack of enhanced speech with residual artifacts. 

The predictive NCSN++M is the \ac{se} model with best speech recognition rates. In spite of the low recognition performance, the generative SGMSE+ obtains the highest non-intrusive quality scores. The hybrid methods strike a middle ground between both approaches. These trends agree with previous works comparing generative and predictive methods~\cite{lemercier2023analysing, deoliveira2023behavior, pirklbauer2023evaluation}. The edge that SE-Mamba gets in POLQA could be attributed to the GAN loss term based on its predecessor PESQ, whose approach shares many similarities. Overall, \ac{asr} does not match the ranking of other metrics directed toward quality and intelligibility.

For the remainder of the paper, we use all of the audio files in EARS-WHAM test for which ground-truths transcriptions are available, totaling 676 samples with \ac{snr} range $[-2.5,17.5]$\,dB.
\vspace{-7pt}
\begin{table}[h]
    \caption{Utterance- and system-level correlations between ASR and human recognition accuracy on the listening experiment data, across all conditions, including clean. The Pearson (PCC) and Spearman's rank (SRCC) correlation coefficients measure linear correlation and monotonic relation, respectively.}
    \label{tab:corr}
    \centering
    \begin{adjustbox}{width=\linewidth}
    \begin{tabular}{lcccc}
        \toprule
         & \multicolumn{2}{c}{Utterance level} & \multicolumn{2}{c}{System level} \\
        \cmidrule(lr){2-3}\cmidrule(lr){4-5}
        ASR Model & PCC & SRCC & PCC \scriptsize{[95\% CI]} & SRCC \scriptsize{[95\% CI]} \\
        \midrule
        QuartzNet 15x5 
        & 0.74 
        & 0.74 
        & 0.74 \scriptsize{[0.58,0.84]}
        & 0.43 \scriptsize{[0.21,0.75]}\\
        wav2vec2 (L) 
        & 0.73 
        & 0.67 
        & 0.79 \scriptsize{[0.64,0.89]}
        & 0.64 \scriptsize{[0.32,0.79]}\\
        Parakeet TDT v2 
        & 0.81 
        & 0.73 
        & 0.93 \scriptsize{[0.82,0.97]}
        & \textbf{1.00} \scriptsize{[0.86,1.00]} \\
        Whisper Base (En) 
        & 0.82 
        & 0.78 
        & \textbf{0.99} \scriptsize{[0.92,0.99]}
        & \textbf{1.00} \scriptsize{[0.75,1.00]} \\
        \hspace{0.2cm} w/o WAcc clipping 
        & 0.32 
        & 0.78 
        & 0.50 \scriptsize{[0.14,0.99]}
        & 0.64 \scriptsize{[0.50,1.00]} \\ 
        Whisper Base 
        & 0.84 
        & \textbf{0.80} 
        & 0.97 \scriptsize{[0.90,0.99]}
        & \textbf{1.00} \scriptsize{[0.82,1.00]} \\
        Whisper (L) v3 Turbo 
        & 0.85 
        & 0.78 
        & 0.97 \scriptsize{[0.90,0.99]}
        & \textbf{1.00} \scriptsize{[0.86,1.00]} \\
        Distil-Whisper (L) v3 
        & \textbf{0.87} 
        & \textbf{0.80} 
        & 0.97 \scriptsize{[0.90,0.99]}
        & 0.96 \scriptsize{[0.86,1.00]} \\
        \bottomrule
    \end{tabular}
    \end{adjustbox}
\end{table}
\vspace{-10pt}
\subsection{Sources of error}
\label{subsec:sdi}

As shown in Equation~\ref{eq:wer}, \ac{wer} combines three types of errors. Representing the total number of words in the reference as $N = S + D + C$, we decompose \ac{wer} into :
\begin{equation}
    \mathrm{WER} = \frac{S + D + I}{S + D + C} = \underbrace{\frac{S}{N}}_{\substack{\text{Substitution}\\\text{Rate}}} + \underbrace{\frac{D}{N}}_{\substack{\text{Deletion}\\\text{Rate}}} + \underbrace{\frac{I}{N}}_{\substack{\text{Insertion}\\\text{Rate}}}
    \label{eq:split}
\end{equation}
These error rates are shown in Fig.~\ref{fig:errors} averaged across all \ac{se} models. We group the samples per input \ac{snr} and present the errors per bin. Most of the errors come in the form of substitutions, likely due to phoneme confusions that change the final word. Whisper models make the least deletions, but the most insertions at lower \acp{snr}, which is in line with studies of hallucinations induced by non-speech~\cite{frieske2024hallucinations, baranski2025investigation}. Due to \emph{looping}, i.e., incorrect repetitions, Whisper incurred \acp{wacc} as low as $-$2061\%.

\subsection{Potential effects on system ranking}

To demonstrate the importance of the \ac{asr} pipeline, we investigate whether some simple variations on the pipeline affect the ranking of the candidate \ac{se} systems. We probe rank consistency via Kendall's $\tau$ across bootstrap samples.
    
\noindent\textbf{Punctuation:} We compare our usual procedure of removing punctuation in the text normalization step with keeping it for \ac{wer} computation. While Parakeet TDT and Whisper output punctuated text, the \ac{ctc} models QuartzNet and wav2vec2 do not and thus experience a severe hit in performance (Approx. $-$10\% \ac{wacc}). Although the degradation is generally consistent across systems, QuartzNet and wav2vec2 experienced changes in ranking in 18.6\% and 16.6\% of the samples, respectively.

\noindent\textbf{Reference text:} So far we have always computed the \ac{wer} / \ac{wacc} using ground truth transcriptions as the reference. However, if the audio in a dataset is not accompanied by transcriptions, one alternative would be to run \ac{asr} on the clean audio target and use its transcription as a reference. Doing so resulted in a general increase of the absolute values by roughly 2\% \ac{wacc}, demonstrating a bias toward the model's own transcriptions. QuartzNet and wav2vec2 experienced changes in ranking in 16.9\% and 18.9\% of the samples, respectively.

\section{Conclusion}

\ac{asr}-based evaluation metrics are often used to evaluate \ac{se} systems. This paper analyzes the behavior of different \ac{asr} models regarding their transcription accuracy scores and subsequent ranking of \ac{se} models. 
Models trained with large-scale, noisy data are the ones that best match the trends in human transcription capabilities, although in our experiments they consistently outperformed humans in absolute numbers. However, scores from these models may hide the effect of residual noise and leverage context, as evidenced by ranking discrepancies with acoustics-focused metrics as ESTOI and POLQA. In some cases, handling of hallucinated outliers also needs to be taken into account. Besides model choice, the exact text processing pipeline can also considerably affect results. In summary, we show that transparently stating and motivating the choice of \ac{asr} model, as well as providing details on the \ac{wer} pipeline are of great importance for the reproducibility, interpretability, and comparability of reported \ac{asr}-based scored in \ac{se} literature.

\ifcameraready

\section{Acknowledgments}

We would like to thank all the test subjects who agreed to take part in the listening experiment. 

\fi

\section{Generative AI Use Disclosure}

No AI writing tools were used for the writing of this paper.

\bibliographystyle{IEEEtran}
\bibliography{mybib}

\end{document}